\documentclass[twocolumn,english,superscriptaddress]{revtex4}
\usepackage[T1]{fontenc}
\usepackage[latin1]{inputenc}
\usepackage{amsmath}
\usepackage{amssymb}

\makeatletter
\usepackage{bm}
\usepackage{braket}

\usepackage{babel}
\makeatother
\begin{document}

\title{Quantum Theory of Orbital Magnetization and its Generalization
  to Interacting Systems}

\author{Junren Shi}

\affiliation{Institute of Physics and ICQS, Chinese Academy of
  Sciences, Beijing 100080, China}

\author{G. Vignale}

\affiliation{Department of Physics, University of Missouri, Columbia,
  MO 65211, USA}

\author{Di Xiao}

\affiliation{Department of Physics, The University of Texas at Austin,
  Austin, TX 78712, USA}

\author{Qian Niu}

\affiliation{Department of Physics, The University of Texas at Austin,
  Austin, TX 78712, USA}

\begin{abstract}
  Based on standard perturbation theory, we present a full quantum
  derivation of the formula for the orbital magnetization in periodic
  systems. The derivation is generally valid for insulators with or
  without a Chern number, for metals at zero or finite temperatures,
  and at weak as well as strong magnetic fields.  The formula is shown
  to be valid in the presence of electron-electron interaction,
  provided the one-electron energies and wave functions are calculated
  self-consistently within the framework of the exact current and spin
  density functional theory.
\end{abstract}
\maketitle

Magnetism is one of the most important properties of materials. Both
spin and orbital motion of electrons can contribute to the total
magnetization.  While the spin magnetization can already be calculated
from first principles with high accuracy by state-of-art methods such
as the spin density functional theory (SDFT), the study of orbital
magnetization is still in a comparatively primitive stage.

A first difficulty arises from the fact that there is still no
theoretically well established formula for calculating the orbital
magnetization of a crystalline solid.  The non-locality of the orbital
magnetization operator $\hat{\bm M}=-e\hat{\bm
  r}\times\hat{\bm\upsilon}$ is the major obstacle to obtaining a
closed formula for an extended periodic system.  Recently, Xiao et
al.~\cite{Xiao2005} and, independently, Thonhauser~{\it et
  al.}~\cite{Thonhauser2005,Ceresoli2006} obtained an orbital
magnetization formula which avoids the non-locality problem and looks
very promising for applications.  However, up to date, there exists no
general quantum mechanical derivation of this formula. The derivation
presented in Ref.~\cite{Xiao2005} relies on the semi-classical
wave-packet dynamics of Bloch
electrons~\cite{Chang1995,Sundaram1999,Xiao2005}, and its validity in
the quantum context is not completely clear.  On the other hand, the
derivation presented in Ref.~\cite{Thonhauser2005} is quantum
mechanical, but relies on the existence of localized Wannier
functions, and cannot be easily generalized to metals or insulators
with non-zero Chern number.  In addition, both derivations are limited
to non-interacting systems.  The shortcomings of these approaches call
for a full quantum mechanical and many-body theory of the orbital
magnetization.
 
A second difficulty is that a first principle calculation of the
orbital magnetization (taking into account many-body effects) should
be based on the spin current density functional theory
(SCDFT)~\cite{Vignale1988}, rather than the conventional SDFT.
Unfortunately, SCDFT has been hindered so far by the lack of reliable
expressions for the magnetization-dependent effective potentials.
This may partly explain why the orbital moments of ferromagnetic
transition metals such as Fe, Co, and Ni calculated in SCDFT were
found to be significantly smaller than the experimentally determined
values~\cite{Ebert1997}.  How problematic these calculations are is
well explained in the review article by Richter~\cite{Richter2001}.
The situation, however, has been rapidly changing in recent years.
The advent of optimized effective
potentials~\cite{Rohra2006,Pittalis2006} which treat exchange exactly
and may systematically be improved for correlations opens new avenues
to the study of magnetic materials.  Applications to atoms and
molecules have already appeared in the literature, and applications to
periodic systems are the obvious next step.

Against this background, the present Letter serves a dual purpose.
First, we present a general derivation of orbital magnetization in
periodic systems based on the standard perturbation theory of quantum
mechanics.  The derivation clarifies the origin of the novel aspects
of the semi-classical derivation, such as the Berry phase correction
to the density of states.  It is generally valid for metals and
insulators with or without a Chern number, at zero or finite
temperatures, in weak or strong magnetic field, and, of course, in the
presence of spin-orbit coupling.  Second, and most important, we
combine this derivation with the exact current and spin density
functional theory~\cite{Vignale1987,Vignale1988}, proving the validity
of the magnetization formula for interacting systems.  We believe that
the magnetization formula, in combination with the recent advances the
construction of optimized effective potential for SCDFT, will turn out
to be a powerful practical tool for the study of systems that have
long defied traditional ab-initio methods.~\cite{Richter2001}

We start from the standard thermodynamic definition of the orbital
magnetization density:
\begin{equation}
  \bm M =-\frac{1}{V}\left(\frac{\partial\Omega}{\partial\bm B}\right)_{T,\mu},
  \label{eq:MT}
\end{equation}
where $\Omega=E-TS-\mu N$ is the grand thermodynamic potential, $V$ is
the total volume of the system, and $\bm B$ is a magnetic field that
only couples to the orbital motion of electrons (but does not
contribute to the Zeeman energy)~\footnote{The separation of spin and
  orbital degrees of freedom is only possible in a semi- relativistic
  framework, such as the one provided by the Pauli equation.  In a
  fully relativistic treatment orbital and spin magnetizations merge
  together, and our formula gives the total magnetization.}.  For
convenience of derivation, we will first calculate the auxiliary
quantity
\begin{equation}
\bm{\tilde M} \equiv-\frac{1}{V}\left(\frac{\partial K}{\partial\bm B}
    \right)_{T,\mu},\label{eq:MK}
\end{equation}
where $K=E-\mu N$. We have $\bm M={\bm{\tilde{M}}} +T\left(\partial
  S/\partial\bm B\right)_{T,\mu}$ and, making use of the Maxwell
relation $\left(\partial S/\partial\bm
  B\right)_{T,\mu}=\left(\partial\bm M/\partial T\right)_{\mu,\bm B}$,
we have the simple relation back to the orbital magnetization:
\begin{equation}
\frac{\partial(\beta\bm M)}{\partial\beta}={\bm{\tilde{M}}} ,
\label{eq:Mt - MK}\end{equation}
where $\beta\equiv1/kT$.

In principle, one can evaluate ${\bm{\tilde{M}}} $ by employing the
standard perturbation theory of quantum mechanics to calculate the
energy correction due to a uniform magnetic field. However, such an
approach will again hit the difficulty of the non-locality of the
orbital magnetization operator. To go around the difficulty, we apply
an external magnetic field that has an infinitely slow spatial
variation~\cite{Vignale1991,Mauri1996}:
\begin{equation}
  \bm B(\bm r)=B\cos(qy)\,\hat{\bm z}
\end{equation}
The slow spatial variation of the field is controlled by the wave
vector $q$, which will tend to zero at the end of the calculation.
The correction to the energy density in this situation can be written,
up to linear order in $ B$, as
\begin{equation}
  \delta K (\bm r) =  -{\bm{\tilde{M}}} \cdot\bm B(\bm r).
\end{equation}
We can then read out ${\bm{\tilde{M}}}$, and take the limit of
$q\rightarrow0$ for a uniform magnetic field. \\
\\

{\it Non-interacting periodic systems} --- For clarity, we first carry
out the perturbation calculation for non-interacting periodic systems.
The single-particle Hamiltonian can be expanded as
$
  \hat{H}  =  \hat{H}_{0}+\hat{V}_{\bm B},
$
where $\hat{H}_{0}$ is the unperturbed Hamiltonian, which yields the
band dispersion $\epsilon_{n\bm k}$ and the corresponding Bloch wave
function $\psi_{n\bm k}(\bm r)=\exp\left(i\bm k\cdot\bm r\right)u_{n\bm
  k}(\bm r)$, and $\hat{V}_{\bm B}$ denotes the coupling to the
external magnetic field:
\begin{equation}
  \hat{V}_{\bm B}  =  \frac{e}{2}\left[\hat{\bm\upsilon}
  \cdot\bm A(\bm r)+\bm A(\bm r)\cdot\hat{\bm\upsilon}\right],
\end{equation}
where $\hat{\bm\upsilon}$ is the velocity operator, and $\bm A(\bm r$)
is the vector potential
\begin{equation}
  \bm A(\bm r)  =  -B\frac{\sin qy}{q}\hat{\bm x},
\end{equation}
which corresponds to the magnetic field discussed earlier.  It is
natural to define the grand-canonical ensemble energy density as
$K(\bm r)=\sum_{n \bm k}f_{n\bm k} \mathrm{Re} \{ \psi_{n\bm
  k}^{*}(\bm r) \hat{K}\psi_{n\bm k}(\bm r) \}$, where
$\hat{K}=\hat{H}-\mu\hat{N}$, and $f_{n\bm k}$
are the occupation number of the single-electron states of band index
$n$ and crystal-momentum $\bm k$~\footnote{The definition here differs
  slightly in the kinetic energy density from the conventional one by
  a total divergence.  Its perturbation by the magnetic field vanishes
  in the limit of $q\to 0$, and therefore does not affect our final
  result.}. To first order in the perturbation, three kinds of terms
arise from changes in the occupation number, the operator $\hat K$,
and the wave function:
\begin{multline}
  \delta K(\bm r)= \mathrm{Re} \sum_{n\bm k}  \left\{ \delta f_{n\bm k}
      \psi_{n\bm k}^{*}\hat{K}_0\psi_{n\bm k} + f_{n\bm k}
       \psi_{n\bm k}^* \hat{V}_{\bm B} \psi_{n\bm k}
    \right. \\
    \left. + f_{n\bm k}\left(\psi_{n\bm
        k}^{*}\hat{K}_{0}\delta\psi_{n\bm k} + \delta \psi_{n\bm
        k}^{*}\hat{K}_{0} \psi_{n\bm k} \right)\right\}.
  \label{eq:deltaK}
\end{multline}
The orbital magnetization can be determined from the appropriate
Fourier component of the energy density, i.e.,
\begin{equation} 
  \tilde{M}_{z}=-\frac{2}{VB}\int\mathrm{d}\bm r\delta K(\bm r)\cos qy.
  \label{eq:MKz}
\end{equation}

It is easy to verify that the first two terms of Eq.~(\ref{eq:deltaK})
does not contribute to $\tilde{M}_z$, and only the contribution from the
change of the wave functions remains.  The first order perturbation to
the wave function reads:
\begin{widetext} 
  \begin{equation}
    \delta\psi_{n\bm k}(\bm r)=-\frac{eB}{4iq}\sum_{n^{\prime}}
    \left\{ \frac{e^{i(\bm k+\bm q)\cdot\bm r}\Ket{u_{n^{\prime}\bm k+\bm q}}
        \Braket{u_{n^{\prime}\bm k+\bm q}|\hat{\upsilon}_{x}(\bm k)+
          \hat{\upsilon}_{x}(\bm k+\bm q)|u_{n\bm k}}}
      {\epsilon_{n\bm k}-\epsilon_{n^{\prime}\bm k+\bm q}}
      -\left(\bm q\rightarrow-\bm q\right)\right\} ,
    \label{eq:delta phi}
  \end{equation}
  where $\hat{\bm\upsilon}(\bm k) \equiv\partial\hat{H}_{0}(\bm
  k)/\partial(\hbar\bm k)$ is the velocity operator, and
  $\hat{H}_{0}(\bm k)$ is defined from the unperturbed Hamiltonian by
  shifting the momentum operator with $\hbar \bm k$.  The transformed
  Hamiltonian acts on the periodic functions, with $u_{n\bm k}$ being
  its eigenfunctions and the band energy $\epsilon_{n\bm k}$ its
  eigenvalues.  Making use Eqs.~(\ref{eq:deltaK}--\ref{eq:delta phi}),
  we have:
  \begin{equation}
    \bm{\tilde{M}}_{z}= 
    \frac{e}{4q}\mathrm{Im}\sum_{nn^{\prime}\bm
      k}\frac{(\epsilon_{n\bm k}+\epsilon_{n^{\prime}\bm k+\bm
        q}-2\mu)\Braket{u_{n\bm k}|u_{n^{\prime}\bm k+\bm
          q}}\Braket{u_{n^{\prime}\bm k+\bm q}|\hat{\upsilon}_{x}(\bm
        k)+\hat{\upsilon}_{x}(\bm k+\bm q)|u_{n\bm k}}}{\epsilon_{n\bm
        k}-\epsilon_{n^{\prime}\bm k+\bm q}}(f_{n\bm
      k}-f_{n^{\prime}\bm k+\bm q}).
  \end{equation} 

  Taking the long-wavelength limit $q\rightarrow0$, we
  obtain:
  \begin{multline}
    \tilde{M}_{z}=\frac{e}{2}\mathrm{Im}\sum_{n\ne n^{\prime}\bm k}
    \frac{(\epsilon_{n\bm k}+\epsilon_{n^{\prime}\bm k}-2\mu)
      \Braket{u_{n\bm k}|\partial u_{n^{\prime}\bm k}/
        \partial k_{y}}\Braket{u_{n^{\prime}\bm k}|
        \hat{\upsilon}_{x}(\bm k)|u_{n\bm k}}}
    {\epsilon_{n\bm k}-\epsilon_{n^{\prime}\bm k}}(f_{n\bm k}-f_{n^{\prime}\bm k})\\
    +e\sum_{n\bm k}(\epsilon_{n\bm
      k}-\mu)\mathrm{Im}\left[\upsilon_{x}^{n}(\bm k)\Braket{u_{n\bm
          k}|\frac{\partial u_{n\bm k}}{\partial
          k_{y}}}+\Braket{\frac{\partial u_{n\bm k}}{\partial
          k_{y}}|\hat{\upsilon}_{x}(\bm k)|u_{n\bm
          k}}+\frac{1}{2}\Braket{u_{n\bm
          k}|\frac{\partial\hat{\upsilon}_{x}(\bm k)}{\partial
          k_{y}}|u_{n\bm k}}\right]f_{n\bm
      k}^{\prime},
    \label{eq:MKz-1}
  \end{multline} 
  where $f^{\prime}_{n\bm k}\equiv\partial
  f(\epsilon_{n\bm k})/\partial\epsilon_{n\bm k}$.  The second term comes from the
  intra-band contribution with $n=n^{\prime}$.  Eq.~(\ref{eq:MKz-1})
  can be further simplified with the help of the relations $
  \Braket{u_{n^{\prime}\bm{k}}|\hat{\upsilon}_{x}(\bm{k})|u_{n\bm{k}}}
  =(1/\hbar)(\epsilon_{n\bm{k}}-\epsilon_{n^{\prime}\bm{k}})
  \Braket{u_{n^{\prime}\bm{k}}|\partial u_{n\bm{k}} / \partial k_x} $
  for $n\ne n^{\prime}$, and $ \mathrm{Im} [\dots ]=
  (1/\hbar)\mathrm{Im} \langle \partial u_{n\bm k}/\partial
    k_{y}|\epsilon_{n}(\bm k)-\hat{H}_{0}(\bm k)|\partial u_{n\bm k} /
    \partial k_{x} \rangle,$ where $\left[\dots\right]$ denotes the
  expression inside the square bracket in Eq.~(\ref{eq:MKz-1}).
  Combining these relations, and generalizing the result to the other
  components of ${\bm{\tilde{M}}} $, we obtain finally:
\begin{equation}
  {\bm{\tilde{M}}} =-\frac{e}{2\hbar}i\sum_{n\bm k}
  \left\{ \Braket{\frac{\partial u_{n\bm k}}{\partial\bm k}|
      \left[\epsilon_{n}(\bm k)+\hat{H}_{0}(\bm k)-2\mu\right]
      \times|\frac{\partial u_{n\bm k}}{\partial\bm k}}
    f_{n\bm k}-(\epsilon_{n\bm k}-\mu)\Braket{\frac{\partial u_{n\bm k}}
      {\partial\bm k}|[\epsilon_{n}(\bm k)-\hat{H}_{0}(\bm k)]
      \times|\frac{\partial u_{n\bm k}}{\partial\bm k}}f_{n\bm k}^{\prime}\right\} 
  .\label{eq:MK-final}\end{equation}
\end{widetext}

The auxiliary and proper orbital magnetization become the same at zero
temperature.  In this case, the second term in the above expression
Eq.~(\ref{eq:MK-final}) vanishes because $f^\prime$ becomes a
$\delta$-function of $(\epsilon_{n\bm k}-\mu)$.  The result is in
perfect agreement with the semiclassical formula of zero temperature
orbital magnetization of Xiao {\it et al.}~\cite{Xiao2005}. For
finite temperatures, we integrate Eq.~(\ref{eq:Mt - MK}) and obtain:
\begin{equation}
\bm M=\sum_{n\bm k}\left\{ \bm m_{n}(\bm k)f_{n\bm k}+\frac{e}
  {\hbar}\bm\Omega_{n}(\bm k)\frac{1}{\beta}\ln\left(1+e^{-\beta(\epsilon_{n\bm k}-\mu)}
  \right)\right\} ,
\label{eq:MT-final}
\end{equation}
where $\bm m_{n}(\bm k) \equiv (e/2\hbar)i\langle\bm\nabla_{\bm
  k}u_{n\bm k}|[\epsilon_{n}(\bm k)-\hat{H}_{0}(\bm
k)]\times|\bm\nabla_{\bm k}u_{n\bm k}\rangle$ is the orbital moment of
state $n, \bm k$ and $\bm\Omega_{n}(\bm k)\equiv
i\Braket{\bm\nabla_{\bm k}u_{n\bm k}|\times|\bm\nabla_{\bm k}u_{n\bm
    k}}$ is the Berry curvature. The same expression was also obtained
in Ref.~\cite{Xiao2006}.

Thus, all the previously known results are recovered by our fully
quantum mechanical formalism.  These results are valid not only for
insulators with or without a Chern number, but also for metals at zero
or finite temperatures. This implies that the semi-classical results
are in general valid to linear order in the external fields. In
hindsight, this should have been expected, because the semiclassical
theory is designed to be exact in the limit of long length scales in
the perturbation to the Hamiltonian.  In our case, this length scale
(through the vector potential) does diverge in the limit of vanishing
magnetic field.  \\

{\it Generalization to interacting systems} --- It is very
desirable to generalize the above results to an interacting system.
This can be done exactly within the framework of the current and spin
density functional theory (CSDFT)~\cite{Vignale1987,Vignale1988}.
CSDFT is a generalization of the spin density functional theory which
includes the current density as an independent variable for the energy
functional and thus provides direct access, via a variational
principle of the Hohenberg-Kohn type, to the current density and the
orbital magnetization of the thermodynamic equilibrium
ensemble~\cite{Ebert1997}. Following the formalism of
Ref.~\cite{Vignale1988}, the many-body problem can be reduced to
solving an effective one-body Schr\"{o}dinger equation:
\begin{equation}
  \left[\frac{1}{2m}\left(-i\hbar\bm\nabla+ e\bm
      A_{\sigma}^{\prime}(\bm r)\right)^{2}+V_{\sigma}^{\prime}(\bm
    r)\right]\psi_{i\sigma}(\bm r)=\epsilon_{i\sigma}\psi(\bm
  r),\label{eq:cslda}
\end{equation} 
with
\begin{align}
  V_{\sigma}^{\prime}= & V_{\sigma}+V_{H}+V_{\mathrm{xc}\sigma}
  +\frac{e^{2}}{2m}\left(\bm A_{\sigma}^{2}-\bm A_{\sigma}^{\prime2}\right),\\
  \bm A_{\sigma}^{\prime}= & \bm A_{\sigma}+\bm
  A_{\mathrm{xc}\sigma}.
\end{align} 
Here $V_{\sigma}$ and $\bm A_{\sigma}$ are the external scalar and
vector potential, respectively, acting on the $\sigma$ component of
the spin ($\sigma=\uparrow\rm or \downarrow$);
$V_{H}=e^{2}\int\mathrm{d}\bm r^{\prime}n(\bm r^{\prime})/|\bm r-\bm
r^{\prime}|$ is the Hartree potential, and $V_{\mathrm{xc}\sigma}$ and
$\bm A_{\mathrm{xc}\sigma}$ are the exchange-correlation (xc) scalar
and vector potentials derived from the xc energy functional
$\Omega_{\mathrm{xc}}[n_{\sigma},\bm j_{p\sigma}]$ according to the
formulas $V_{\mathrm{xc}\sigma}=\delta \Omega_{\mathrm{xc}}/\delta
n_{\sigma}$, $e\bm A_{\mathrm{xc}\sigma}=\delta
\Omega_{\mathrm{xc}}/\delta\bm j_{p\sigma}$. The density,
$n_{\sigma}(\bm r)$, and the paramagnetic current density, $\bm
j_{p\sigma}(\bm r)$ are to be determined self-consistently from the
solutions of the above equation according to the formulas
$n_{\sigma}(\bm r)=\sum_{i}|\psi_{i\sigma}(\bm
r)|^{2}f(\epsilon_{i\sigma})$ and $\bm j_{p\sigma}(\bm
r)=(-i\hbar/2m)\sum_{i}[\psi_{i\sigma}^{*}(\bm
r)\bm\nabla\psi_{i\sigma}(\bm r)-\bm\nabla\psi_{i\sigma}^{*}(\bm
r)\psi_{i\sigma}(\bm r)]f(\epsilon_{i\sigma})$.  At finite
temperature, the thermodynamic potential functional can be written
as~\cite{Vignale1988}:
\begin{multline}
  \Omega=-\frac{1}{\beta}\sum_{i\sigma}\ln
  \left[1+e^{-\beta(\epsilon_{i\sigma}-\mu)}\right]\\
  -\frac{1}{2}e^{2}\int\int\mathrm{d}\bm r
  \mathrm{d}\bm r^{\prime}\frac{n(\bm r)n(\bm r^{\prime})}{|\bm r-\bm r^{\prime}|}
  -\sum_{\sigma}\int\mathrm{d}\bm rn_{\sigma}(\bm r)V_{\mathrm{xc}\sigma}(\bm r)\\
  -e\sum_{\sigma}\int\mathrm{d}\bm r\bm j_{p\sigma}(\bm
  r)\cdot\bm A_{\mathrm{xc}\sigma}(\bm
  r)+\Omega_{\mathrm{xc}}[n_{\sigma},\bm j_{p\sigma}],
  \label{eq:K functional}
\end{multline} 
which is a functional of four fields: the densities $n_{\sigma}$ and
$\bm j_{p\sigma}$, and the external potentials $V_{\sigma}$ and $\bm
A_{\sigma}$ -- the last two enter the expression through the
eigenvalues $\epsilon_{i\sigma}$.

To calculate the orbital magnetization one needs to evaluate the
variation $\delta\Omega$ of the thermodynamic potential in response to
a variation of the external magnetic field, which, in turn, is
generated by a variation in the external vector potentials $\delta \bm
A_{\sigma}$. In general, $\delta\Omega$ can be separated into two
contributions: the primary one ($\delta\Omega|_{n_{\sigma},\bm
  j_{p\sigma}}$) arises directly from the variation of the vector
potentials, keeping $n_{\sigma}$ and $\bm j_{p\sigma}$ constant at
their unperturbed values; the secondary one
($\delta\Omega|_{V_{\sigma,\bm A_{\sigma}}})$ might arise from the
changes of $n_{\sigma}$ and $\bm j_{p\sigma}$ at constant external
potentials (these changes would affect $\Omega$ via the modification
of the effective potentials $V_{H}$, $V_{\mathrm{xc}\sigma}$ and $\bm
A_{\mathrm{xc}\sigma}$):
\begin{equation}
  \delta\Omega=\left.\delta\Omega\right|_{n_{\sigma},\bm j_{p\sigma}}
  +\left.\delta\Omega\right|_{V_{\sigma},\bm A_{\sigma}}.
\end{equation}

It is easy to see that $\delta\Omega|_{n_{\sigma},\bm j_{p\sigma}}$
contributes an orbital magnetization that is exactly given by
Eq.~(\ref{eq:MT-final}), as if the system were a noninteracting system
with eigenfunctions and eigenvalues determined by
Eq.~(\ref{eq:cslda}).  This is because the variation of the external
vector potential affects only the eigenvalues $\epsilon_{i\sigma}$ in
the one-body term of Eq.~(\ref{eq:K functional}). To evaluate
$\delta\Omega|_{V_{\sigma},\bm A_{\sigma}}$ it is sufficient to
observe that for given external potentials the thermodynamic potential
$\Omega$ is stationary against small changes of the density and the
current about their equilibrium values: $\delta\Omega/\delta
n_{\sigma}=\delta\Omega/\delta\bm j_{p\sigma}=0$.

Thus we have 
\begin{equation}
  \left.\delta\Omega\right|_{V_{\sigma},\bm A_{\sigma}}=0.
\end{equation}
We then conclude that in the context of the CSDFT, we can treat the
system as an effective one-body system, and use
Eq.~(\ref{eq:MT-final}) to calculate the orbital magnetization, albeit
using the dispersion and wave-functions derived from
Eq.~(\ref{eq:cslda}).  

We stress that this conclusion would not hold true if
the one-electron orbitals and their energies were calculated within
the framework of the {\it ordinary} density functional
theory~\cite{Grayce1994} (as opposed to CSDFT).  In such a formulation
the xc energy functional would be a functional of density {\it and}
magnetic field: $\Omega_{\mathrm{xc}}[n_\sigma,{\bm B}]$.  Then the formula for
the orbital magnetization would include a functional derivative of
$\Omega_{\mathrm{xc}}$ with respect to ${\bm B}$, at variance to the
one-electron formula of Eq.~(\ref{eq:MT-final}).\\

{\it Finite fields} --- Our formula can be applied rigorously to
finite magnetic fields if these fields are rational in the sense that
fluxes through the faces of a unit cell are fractional multiples of
the flux quantum $h/e$.  In this case, one can define Bloch like
eigen-states with respect to magnetic translations, which are ordinary
translations on the crystal combined with gauge transformations.
Orbital magnetization at such a field can then be calculated
perturbatively by adding a small change $\delta B$ to this field.
Both the semiclassical theory and quantum perturbation with respect to
$\delta B$ give the same expression for the orbital magnetization,
provided we use the magnetic Bloch wavefunctions as a basis.  Like
wise, the CSDFT can also be formulated for systems with periodic
boundary conditions with respect to magnetic translations, so the
justification of our results for interacting systems is
straightforward~\cite{Vignale1988}.  Indeed, current density
functional theory has been used to study the formation of an electron
crystal (Wigner crystal) at very high magnetic
field~\cite{Vignale1993}.

The situation for irrational fields is a bit tricky if one insists on
rigorous results \cite{Gat2003}.  One may consider an irrational field as
a limit of sequence of rational fields.  This is possible if the 
magnetization depends on the field continuously, which is expected 
to be the case when the temperature is finite.  Indeed, when the 
fast de Haas-van Alphen oscillations are smeared out, the average
magnetization changes continuously with the Fermi energy~\cite{Gat2003}.  
For a fixed Fermi energy, the average magnetization also changes 
continuously with the magnetic field.   Therefore, we expect that the 
average magnetization is a continuous function of magnetic field at 
a fixed density of electrons.\\

In summary, we have presented a full quantum derivation of the orbital
magnetization formula. The derivation is generally valid for
insulators with or without a Chern number, for metals at zero or
finite temperatures, and at weak as well as strong magnetic fields.
We also find that the resulting formula is directly applicable to
interacting systems provided one uses one-electron energies and wave
functions obtained from the self-consistent solution of the Kohn-Sham
equation of current and spin-density functional theory.

This work was supported by National Science Foundation of China
No.~10604063. GV was supported by NSF (DMR-031368), DX was supported
by NSF (DMR-0404252/0606485), and QN by the Welch Foundation and DOE
(DE-FG03-02ER45958).


\end{document}